# Title: Designing metainterfaces with specified friction laws

**Authors:** Antoine Aymard[1], Emilie Delplanque[1], Davy Dalmas[1], Julien Scheibert[1]*

**Affiliations:**

[1]Univ Lyon, CNRS, Ecole Centrale de Lyon, ENTPE, LTDS, UMR5513; Ecully, 69130, France.

*Corresponding author. Email: julien.scheibert@cnrs.fr

**Abstract:** Many devices, including touchscreens and robotic hands, involve frictional contacts. Optimizing those devices requires fine control of the interface's friction law. We lack systematic methods to create dry contact interfaces whose frictional behaviour satisfies preset specifications. We propose a generic surface design strategy to prepare dry rough interfaces that have predefined relationships between normal and friction forces. Such metainterfaces circumvent the usual multiscale challenge of tribology, by considering simplified surface topographies as assemblies of spherical asperities. Optimizing the individual asperities' heights enables specific friction laws to be targeted. Through various centimeter-scaled elastomer-glass metainterfaces, we illustrate three types of achievable friction laws, including linear laws with a specified friction coefficient and unusual non-linear laws. This design strategy represents a scale- and material-independent, chemical-free pathway toward energy-saving and adaptable smart interfaces.

**One-Sentence Summary:** Multicontact interfaces are designed to provide on-demand friction coefficients and meet unnatural friction specifications.

**Main Text:**

Despite centuries of investigations, a comprehensive understanding of friction is still lacking. For instance, predicting from first principles the value of the friction force, $F$, of a given dry contact interface remains out of reach, mainly because of the multiscale character of surfaces and the multi-physics nature of contact interactions (*1*). Thus, time- and resource-consuming experimental tests remain necessary to calibrate the frictional behavior of a contact interface as soon as any change is brought to the material pair, shape of the solids, loading conditions, surface finish or environmental conditions. This inability to master friction is a major obstacle to the optimization of devices whose function relies on dry contact interfaces. For soft interfaces, these devices include sport accessories [*e.g.,* racket coatings (*2*), shoe soles (*3*, *4*)], robotic grasping devices (*5–7*), haptic feedback tools for virtual reality (*8*), and conveyor belts (*9*).

At the present time, surface functionalization is the main approach that is followed worldwide to provide contact interfaces with improved frictional capabilities (*1*, *10*). It often consists of creating a certain surface topography at various length scales or adding a homogeneous or heterogeneous thin coating at the solid surface. Unfortunately, despite many successes in a variety of specific cases (*11–15*), this approach is still based on trial and error. It does not offer a general, systematic design strategy to convert a set of frictional specifications into an actual interface that offers the expected behavior. By circumventing the main pitfalls that make it difficult to understand friction in natural interfaces, we propose and validate a general design strategy to prepare multicontact interfaces with on-demand frictional features.

## RESULTS

### General interface design strategy

To bypass the multiscale and multiphysics challenges of friction along rough interfaces, we propose a compromise between simplicity and richness in the surface description, by considering flat-flat contact interfaces between a smooth and a rough surface. Once the material pair is fixed, the designable feature of the interface is the topography of the rough surface, which is built as an ensemble of individual microasperities, with both well-controlled geometrical properties and calibrated contact and friction behaviors against the smooth counter surface. The richness of the emerging macroscale behavior stems from the countless possible combinations of geometrical properties of all individual asperities.

Just as the microstructures of the materials can be engineered to provide metamaterials with macroscale properties that are rarely found in nature [see (*16–19*) for mechanical metamaterials], we propose a design strategy (Fig. 1) to prepare contact interfaces with complex predefined frictional behavior. We denote these as metainterfaces. The strategy starts with a target friction law expressed as the macroscopic relationship $F_{\text{target}}(P)$ between the normal load $P$ applied to the interface and the target macroscopic friction force $F_{\text{target}}$. This law is the input of an inversion step, the output of which is a geometrical description of the surface's topography, including the number of necessary asperities and the list of their individual properties (shape, size, height, position). The inversion is based on two main ingredients. First, the indentation and frictional behaviors of a single microcontact are obtained through a preliminary calibration (top-right illustration in Fig. 1) which may, but need not, be captured by an existing tribological model. Crucially, those calibrated behaviors contain any specific effect due to the manufacturing process, interface physico-chemistry or surface contamination. Second, a suitable asperity-based friction model, able to predict the global frictional behavior as the collective response of the population of asperities, is identified. Depending on the expected relevant physics, the model can range from analytical (*20–22*) to numerical (*23*), through artificial intelligence [*e.g.,* by extending the approaches of previous work (*24*, *25*) to asperity-based descriptions of surface topography]. Based on the inverted asperities' geometries, a corresponding sample can be manufactured. The material, and characteristic size of the prescribed asperities contribute to the selection of the relevant manufacturing method. Finally, shearing tests against the smooth counter surface enable determination of the resulting friction law, $F(P)$, and, through direct comparison with $F_{target}(P)$, assessment of the overall reliability of the workflow. Discrepancies with respect to the target friction law may arise from an incomplete calibration of the single asperity behavior, incorrect assumptions in the friction model, or manufacturing imperfections.

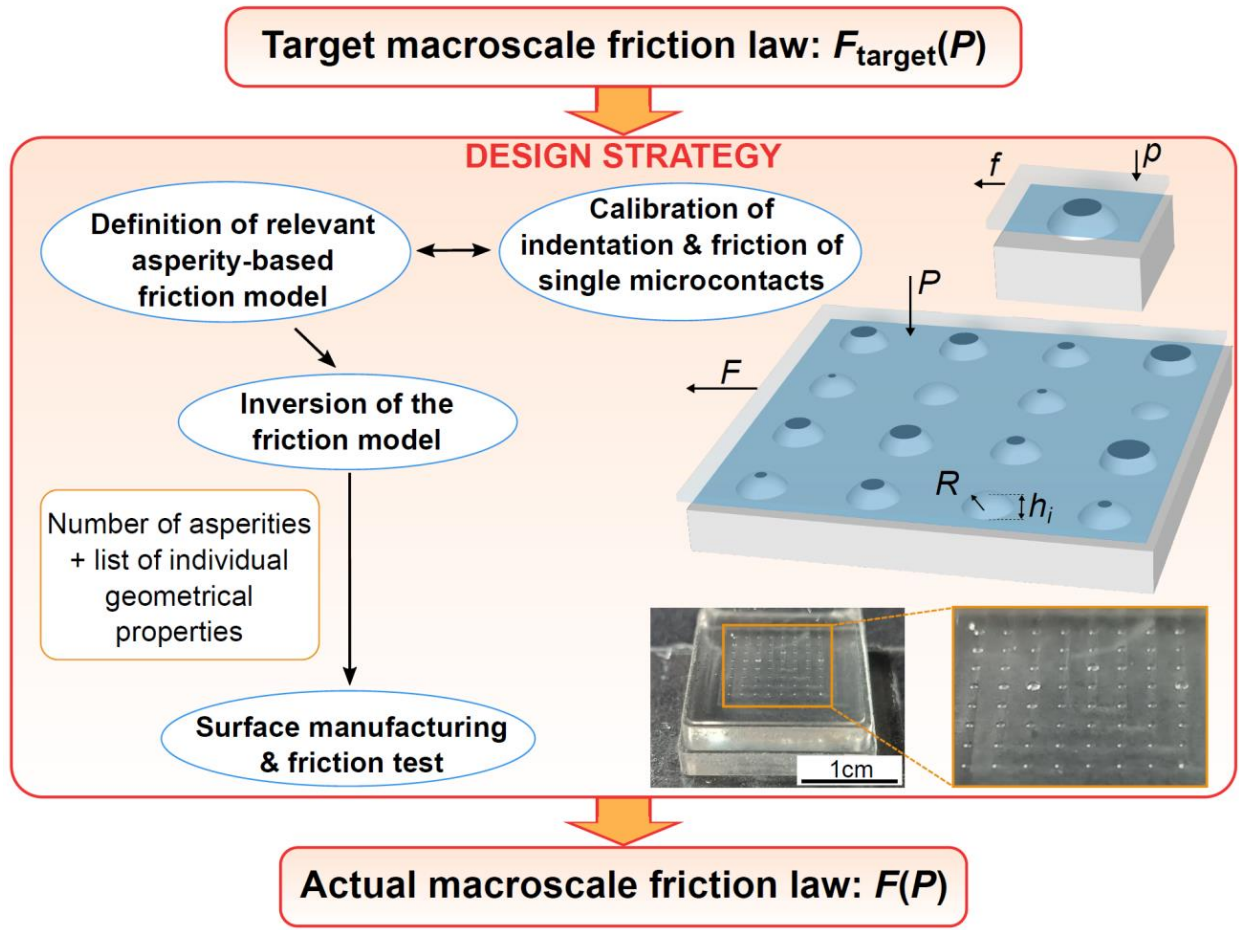

**Fig. 1. Flowchart of the design strategy**. Shown at the top right is an illustration of a single microcontact submitted to a normal force $p$ and a friction force $f$. An illustration of a metainterface submitted to normal and friction forces, $P$ and $F$, is shown at the middle right. Dark gray ellipses represent real contact regions. The topography is made of $N$ asperities, each with specifically designed geometrical properties (here spherical caps of height $h_i$ and curvature radius $R$). At the bottom right is an image of a typical centimetric elastomer-based realization of such a textured sample. Photographs of metainterfaces can be found in Figs. 3 to 5.

**Applying the strategy to centimetric elastomer/glass interfaces**

Although the generic flowchart shown in Fig. 1 is expected to be applicable to a large variety of frictional interfaces, the following scientific and technological choices represent only one example of how our design strategy can be implemented. First, we chose to work with contacts between polydimethylsiloxane (PDMS) and glass (*26*), a widely used pair of materials in contact mechanics and friction studies (*1,11*). The behavior of sheared PDMS-glass, single sphere-plane contacts has been extensively characterized recently (*27–30*), which provides useful insights into the single-asperity laws that should be used in the friction model. In particular, the friction force, $F$, is found to be proportional to the contact area, $A$, through $F=\sigma A$, where $\sigma$ is the frictional strength of the PDMS-glass interface. Predicting the macroscale friction force $F$ is thus substantially simplified because, for a given $\sigma$, predicting $F$ reduces to predicting the total contact area $A$ of all microcontacts.

Second, as illustrated in Fig. 1, we chose to use a square lattice of 64 asperities as spherical caps, with a common radius of curvature $R=526\pm5$ μm and distributed summit heights $h_i$. We obtained them by molding a PDMS slab onto an aluminum mold prepared with deterministic spherical holes using a sphere-ended cutting tool in a micro milling machine (*26*) (for a typical image of a PDMS textured sample, see Fig. 1). This method of preparing populations of spherical PDMS asperities has been successfully applied in the literature (*31–33*), but not used to design interfaces with predefined friction laws, as is done in this work. We have calibrated the indentation and shear behavior of single such microcontacts (*26*). Their normal indentation is well-captured by the classical Hertz model of a linear-elastic sphere-rigid plane contact (*34*) (Fig. 2B), with a composite elastic modulus $E^*$ ($E^* = \frac{E}{1-\nu^2}$, where $E$ and $\nu$ are the Young's modulus and Poisson's ratio of the PDMS, respectively). Consistent with previous work (*27–30*), the contact region is initially circular with an area $a_0$ but decreases anisotropically under shear, by ~10 to 15%, down to an area $a_f$ (Fig. 2A) at the onset of sliding (static friction). The ratio $B=a_f/a_0$ thus represents the fraction of the initial contact area that remains when the contact is sheared and just about to slide. We found $B$ to be independent of $a_0$ (Fig. 2C). We also confirmed the proportionality between the contact area and static friction force, $f=\sigma a_f$ (Fig. 2D).

The third choice we made was the type of friction model to be inverted. For the present proof of concept, we found that a very simple linear-elastic asperity-based model was sufficient. We considered $N$ spherical linear-elastic asperities with common curvature radius, $R$, and composite elastic modulus, $E^*$, but distributed summit heights, $h_i$. This population of asperities is brought into contact with a rigid smooth plane, forming a number of microcontacts, which are each assumed to obey both Hertz's model (*34*) under pure compression, and the calibrated friction behavior under shear. Microcontacts are assumed to be independent [see (*26*) for justification], such that their individual in-plane locations become irrelevant. The area of the $i$th microcontact is expressed as $a_{0,i} = \pi R(h_i - \delta)$ if $h_i \geq \delta$ and $a_{0,i}=0$ if $h_i<\delta$, where $\delta$ is the separation between the rigid indenting plane and the base plane that supports the asperities. The macroscopic quantities are simple sums: $A_0 = \sum_{i=1}^{N} a_{0,i}$, $A_f = \sum_{i=1}^{N} a_{f,i}$, $P = \sum_{i=1}^{N} \frac{4E^*}{3\pi^{3/2}R} a_{0,i}^{3/2}$ (Hertz's model), and $F = \sum_{i=1}^{N} f_i = \sigma A_f$, where $A_f = BA_0$. The model parameter values were experimentally calibrated (*26*). Inverting the above friction model consists in determining a suitable list of heights $h_i$ such that the predicted friction law satisfies specifications set a priori.

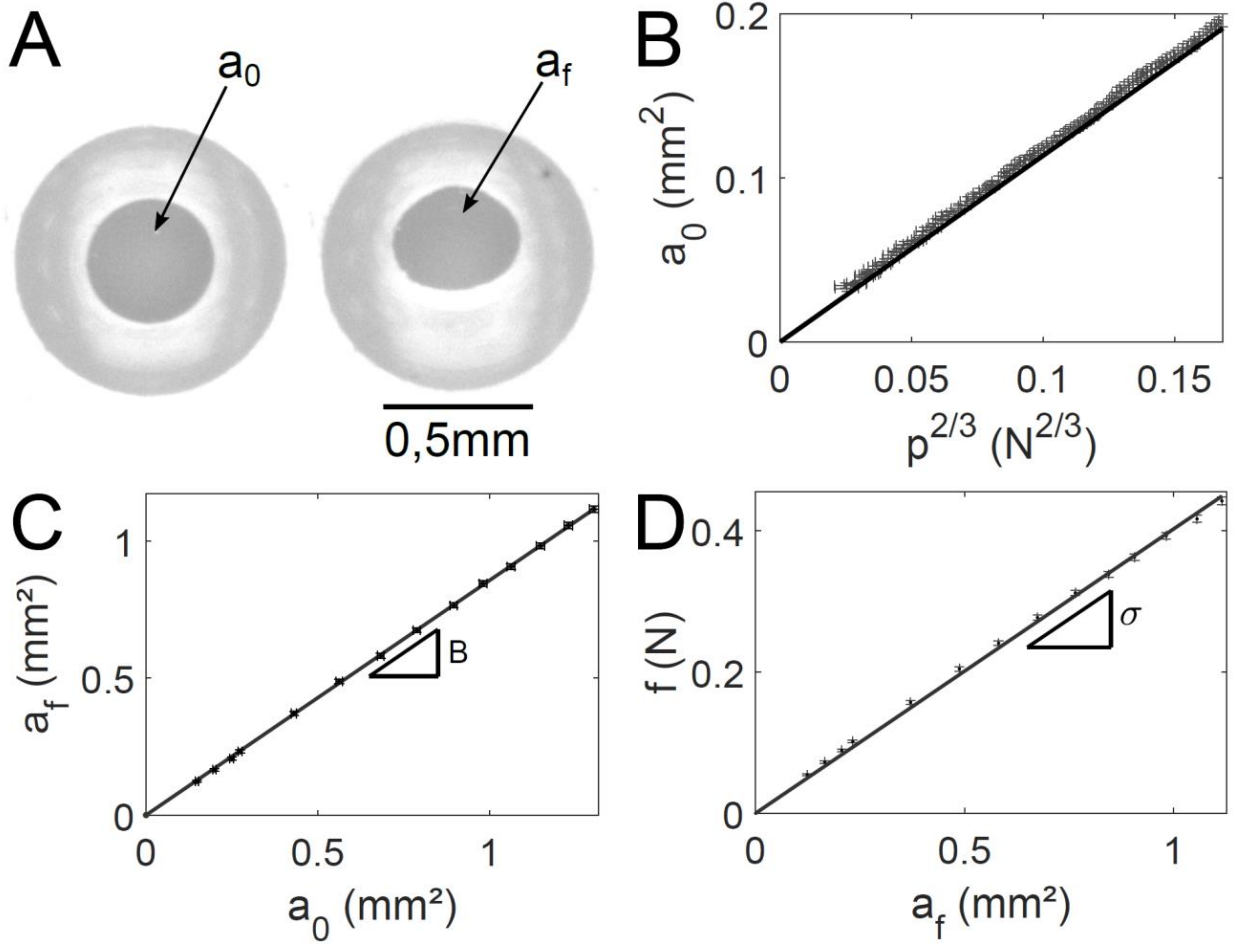

**Fig. 2. Tribological calibration of single microcontacts**. Illustrations for PDMS batch 1. (**A**) Images of a microcontact under normal load $p$=0.061 N with no shear, initial area $a_0$ (dark central region) (left) and at the onset of sliding, area $a_f$ (tangential force applied upward in the image plane) (right). (**B**) $a_0$ versus $p^{2/3}$, for three indentation experiments using three different samples (data points). The solid line is Hertz's prediction $a_0 = \pi \left(\frac{3Rp}{4E^*}\right)^{2/3}$ for $R$=526 µm and $E^*$=1.36 MPa [see (*26*) for information about how those values were determined]. (**C**) Typical evolution of $a_f$ as a function of $a_0$ (data points). The solid line is the linear fit showing the existence of a constant area reduction ratio $B$=$a_f$/$a_0$. (**D**) Typical friction force $f$ versus $a_f$ (data points). The solid line is the linear fit showing the existence of a constant friction strength $\sigma$=$f$/$a_f$. Error bars are ±5 mN for $p$, ±(1 mN+0.01$f$) for $f$, and $\pm 2.8\sqrt{\pi a} \times 10^{-6}$ m² for $a_0$ and $a_f$.

Interestingly, in the model, the friction law $F(P)$ can be rescaled as $\frac{F}{\sigma B}$ versus $\frac{P}{E^*}$, where the function relating $\frac{F}{\sigma B}$ and $\frac{P}{E^*}$ depends only on the geometrical parameters of the asperities, $R$ and $h_i$, and not on the material parameters, $\sigma$, $B$ and $E^*$. Thus, in the following three examples, we illustrate the success of our design strategy using the material-independent relationship $\frac{F}{\sigma B}\left(\frac{P}{E^*}\right)$. This rescaled friction law characterizes solely the effect of surface topography, which is the actual interfacial quantity on which the design is performed.

**Building a friction law from operating points**

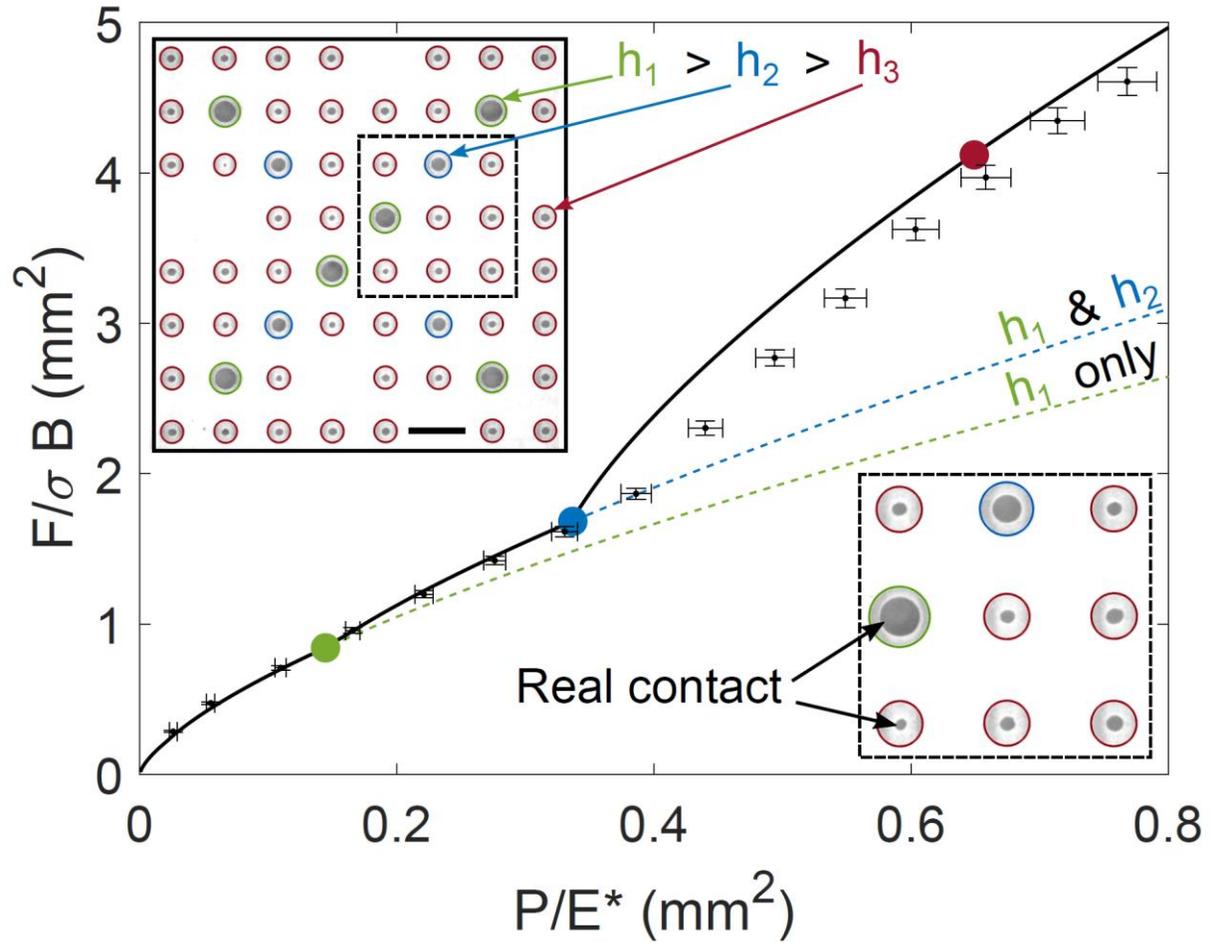

**Fig. 3. Designing an interface that reaches three preset frictional operating points.** The green, blue, and red disks represent targeted operating points. The black points represent measured friction law $\frac{F}{\sigma B}$ versus $\frac{P}{E^*}$. Error bars are calculated as described in (26). The solid line is the model prediction. Dashed lines are continuations of the two first branches of the solid line if the subsequent height levels of the asperities had not been populated. The top-left inset shows a full image of the unsheared contact under $P=0.92$ N. Asperities at each of the three height levels ($h_1>h_2>h_3$) are circled with the same color as their associated operating point (main panel). Scale bar is 1.5 mm. The bottom-right inset shows a magnified view of the real contact of nine microcontacts (the region outlined by the dashed line in the top-left inset).

For our first example, we consider specifications in terms of a list of operating points $\left(\frac{P_i}{E^*}, \frac{F_i}{\sigma B}\right)$ through which the friction law $\frac{F}{\sigma B}\left(\frac{P}{E^*}\right)$ must pass. An infinity of suitable lists of asperity heights $h_i$ exist as the solution to this problem, each giving a different shape of the friction law between operating points. For pedagogical purposes, we adopted an inversion strategy (26) in which each operating point is reached using a single level of asperity height (fig. S1). To pass from the first (trivial) operating point (0,0) to the next, one evaluates the maximum number of asperities with identical heights (first height level) required to approach the operating point from below and adds a single adjustment asperity whose height is selected to reach exactly the desired operating point. The procedure is repeated as many times as the number of additional operating points. As a result, upon normal loading, a prescribed number of new asperities enter the contact as soon as the previous operating point is reached. We applied this procedure to design a metainterface with a predicted three-branched friction behavior passing through three non-aligned operating points (Fig. 3). Our measurement points delineate a friction law that closely approaches the three target friction forces, to better than 5% (2.5% for the first two points). This agreement shows that the design strategy can be used successfully to prepare real-life metainterfaces that target a non-trivial list of friction forces.

The latter design strategy can be extended to an arbitrary number of operating points (although sufficiently smaller than the number $N$ of available asperities), opening the way to the design of interfaces with friction laws defined by many operating points. We expect that inversion procedures that are more versatile than the pedagogical one we present will be developed, for instance, those that invert not only $h_i$ but also other geometrical quantities such as the curvature radius of each asperity.

**Specifying the friction coefficient at a constant material pair**

For our second example, we acknowledged that the most classical descriptor of the frictional properties of an interface is the value of its friction coefficient, µ. Commonly considered to be a characteristic of a pair of materials in contact, the friction coefficient does not relate to a particular operating point but rather to the global shape of the friction law. Indeed, unlike the curve shown in Fig. 3, in most natural or human-made interfaces, the friction law $F(P)$ is found to be linear [Amontons-Coulomb friction (*1*, *35*)], with µ being the slope of the law. Hence, we targeted linear friction laws with tunable slopes.

For surface topographies made of spherical asperities that obey Hertz's indentation law, as in our metainterfaces, a statistical framework relating the contact area $A_0$ to the normal load $P$ exists [see Greenwood and Williamson's model (*20*)]. This framework predicts a linear $A_0(P)$ relationship when the asperities' heights, $h_i$, follow a probability distribution that is exponential ($e^{-h/\lambda}/\lambda$, where $\lambda$ is the scale parameter of the exponential). We adapted this model to account for the existence of a maximum asperity height $h_m$ due to the finite number of asperities, and extended it to predict linear-like friction laws [details in (*26*)]. In particular, we derived the value of the slope of $\frac{F}{\sigma B}\left(\frac{P}{E^*}\right)$ for large $P$: $m = \sqrt{\frac{\pi R}{\lambda}} \frac{1-e^{-x}}{\mathrm{erf}(\sqrt{x}) - \frac{2}{\sqrt{\pi}}\sqrt{x}e^{-x}}$, where $x = h_m/\lambda$. $m$ is a rescaled estimator for the friction coefficient of the interface ($m \approx \frac{E^*}{\sigma B}\mu$).

For illustrative purposes, we fixed $R$ and $h_m$ and we targeted two different slopes $m$. Using the above expression of $m$, we identified the two $\lambda$ that were suitable to generate two metainterfaces, each with one of the target slopes (*26*). The two friction laws that we obtained are linear and their slopes match the predictions to better than 2.1% (Fig. 4). This agreement validates that our design strategy quantitatively enables the prescription of the slope of a linear friction law [after a nonlinear part at small $P$ that is due to the finite value of $h_m/\lambda$ and is responsible for a nonvanishing intercept; see (*26*)]. The slopes $m$ that we achieved for the rescaled friction law $\frac{F}{\sigma B}\left(\frac{P}{E^*}\right)$ differ by a ratio of 1.42 (1.60 for the friction coefficients $\mu \approx \frac{\sigma B}{E^*}m$). For a discussion about the range of achievable friction coefficients, see (26). We emphasize that the results provide a practical way to tune the friction coefficient of a contact interface by manipulating the surface topography only, that is, without bringing any change to the bulk materials or physicochemistry of their surfaces.

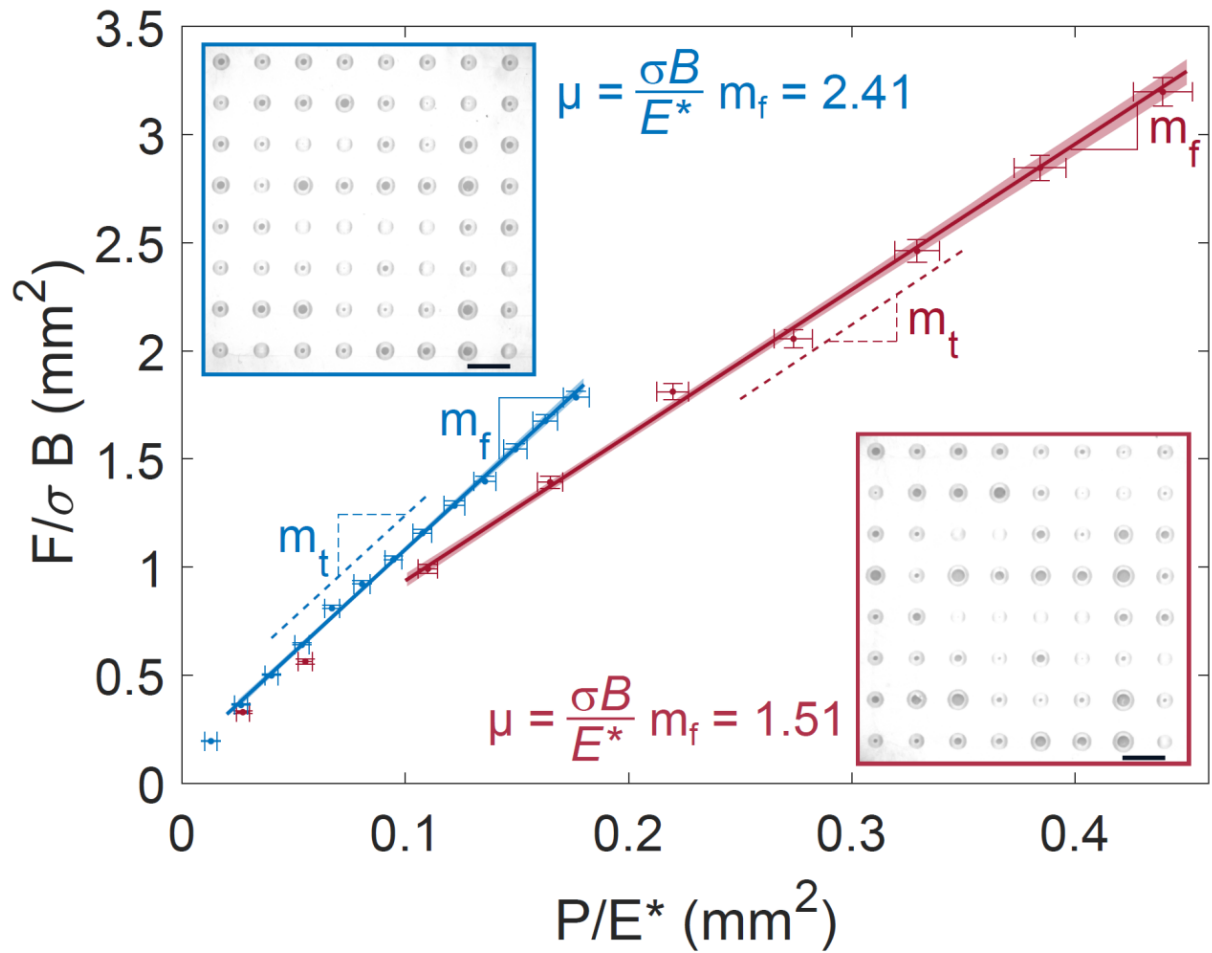

**Fig. 4. Specifying the friction coefficient at a constant material pair.** Experimental friction laws (data points) of two metainterfaces with exponential-like distributions of asperity heights (see table S1 and (26)). Error bars are as described in Fig. 3. Lines are linear fits of the data [over the last 12 (7) points for the blue (red) data]. The red and blue shaded regions around the lines represent the 68% confidence interval on the linear regression. The fitted slopes $m_f$ of the rescaled laws, 6.73±0.22 (red) and 9.53±0.25 (blue) [which correspond to friction coefficients $\mu \approx \frac{\sigma B}{E^*} m_f$=1.51 and 2.41, typical of PDMS-glass rough contacts (27, 29)), match the targeted slopes (dashed lines) $m_t$=6.87 (red) and 9.43 (blue), to better than 2.1%. The insets show typical photographs of the two metainterfaces under pure normal force $P$=0.39 N (blue) or 0.93 N (red). Scale bars are 1.5 mm.

**Building friction laws with multiple specified linear branches**

For our third example, we show that unnatural, non-linear-shaped friction laws can also be designed, enabling one to go beyond Amontons-Coulomb-like friction. As an illustration, we targeted a succession of two linear branches with specified slopes, $m_1$ and $m_2$, and a crossover at a specified critical normal load, $P_c/E^*$. We first describe how such a friction law can be obtained [see (26)], using a weighted exponential distribution of heights, which generates two sub-populations of asperities. We then explain how to invert the law and find a suitable list of asperity heights, $h_i$ [see (26)].

We demonstrate that such bilinear-shaped friction laws can indeed be achieved (Fig. 5) through two different realizations. As targeted, both realizations share the same two slopes, which are different by a ratio of 1.75, but have two different cross-over normal loads, $P_c/E^*$ (ratio 1.94). The slopes values are fulfilled by better than 5.2% (10% for the ratio of $P_c/E^*$), further demonstrating experimentally the success of our design strategy in satisfying quantitatively non-trivial friction prescriptions.

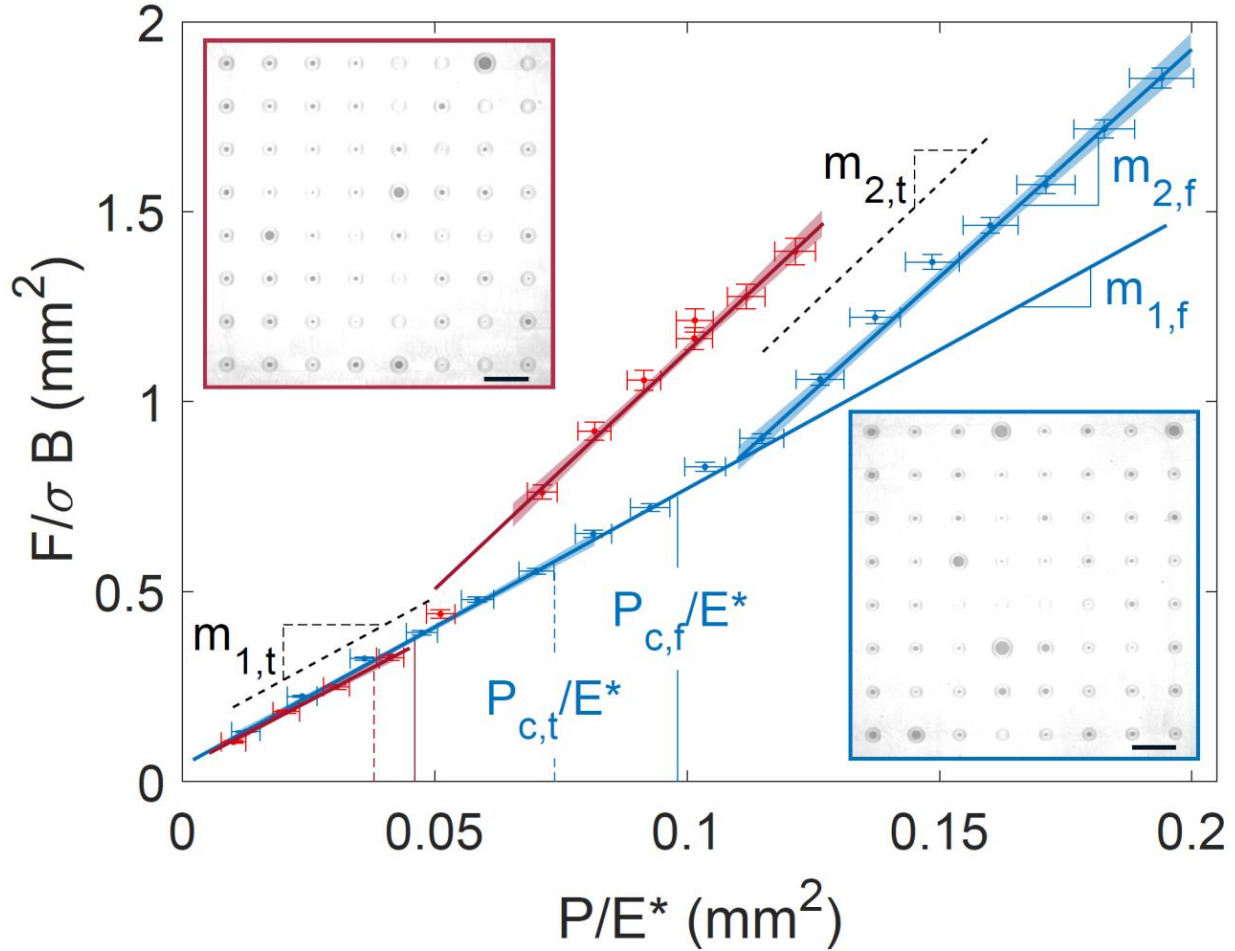

**Fig. 5. Friction law made of two specified linear branches.** Two experimental realizations of bilinear friction laws (data points), based on height distributions given by eq. 6 (table S2). Error bars are as described in Figs. 3 and 4. Solid lines are linear fits of each branch [six (four) first data points for the first branch of the blue (red) curve; eight (seven) last data points for the second branch of the blue (red) curve]. The fitted slopes $m_f$ of both linear-like branches in each curve match the target slopes $m_t$ (dashed black lines), 7.15 and 12.52, to better than 5.2%. Vertical lines show target ($P_{c,t}/E^*$, dashed lines) and observed ($P_{c,f}/E^*$, solid lines) crossover normal loads (in the same colors as the corresponding curves). Insets show typical photographs of the metainterfaces that underlie both friction laws under pure normal force $P$=0.23 N (red) or 0.34 N (blue). The frame color refers to that of the corresponding friction law. Scale bars are 1.5 mm. See movie S1 for a qualitative illustration experiment on analogous metainterfaces.

**DISCUSSION**

Understanding friction in its generality remains a formidable challenge and has been for centuries. Thus, controlling friction based on generic principles is often considered unreachable. Our work demonstrates that the friction of dry elastomer-based metainterfaces can actually be finely tuned, when the surface topography is designed according to our proposed strategy (Fig. 1). We emphasize that our three examples far from exhaust the myriad possible variations around our design strategy, which provides a simple framework that enables friction-control opportunities, starting from, but not limited to, the universally used friction coefficient. Such topography-based control confirms in particular that the static friction coefficient is not a constant for a given pair of materials, as predicted by models (*22, 23, 36*) and observed in stick-slip experiments (*37*).

The potentially accessible friction laws $F(P)$ are countless, in terms of both their nonlinear shape and the way specifications are expressed (forces, slopes, etc.). Those alternative, richer prescriptions likely require the inversion of more topographic features than the individual asperities' heights, including their individual radii of curvature and in-plane locations. Such additional degrees of freedom should prompt the development and use of advanced friction models (*22, 38*) and inversion tools (*25, 39*) that are not limited to the analytical approaches proposed here as pedagogical illustrations.

Static friction is not the only possible target functionality, because the design strategy can presumably be adapted to control other topography-related quantities, including interfacial stiffnesses and adhesion forces. Such extensions require different types of calibrations of individual microcontacts that are dedicated to those target properties. In this respect, note that in our experiments, the friction design strategy can be applied indifferently to the static friction force (as in Figs. 3 to 5) or sliding friction force [see (*26*)].

Our design strategy should be applicable to other materials pairs than PDMS/glass. For materials whose contact remains elastic (see (*26*) for a criterion), the approach developed in our examples should hold, with material changes being accounted for by the rescaled, material-independent version of the friction law, $\frac{F}{\sigma B}\left(\frac{P}{E^*}\right)$. For non-elastic materials, qualitative changes might need to be brought to the calibrated microcontact behavior laws, friction model, and related inversions. For instance, for elastoplastic contacts, enriched asperity-based friction models are required (*40, 41*). In addition, the plasticity-induced irreversible modifications of the topography will likely affect all successive uses of the metainterface such that the friction law is no longer a constant but rather a loading-history-dependent function.

Irreversible changes of the topography may also arise as a result of wear-induced loss of material. Upon sliding, various wear mechanisms may occur, including progressive material removal that leads to blunted asperities, and asperity-sized debris formation due to fracture processes (*42*). With such alterations of the surface topography, the behavior of metainterfaces will progressively deviate from the target one. In this context, evaluating the lifetime of metainterfaces as a function of the materials, roughness and loading conditions will be an important step before use in any specific engineering application.

We expect our design strategy to be applicable over a large variety of length-scales. We used asperities with a sub-millimetric radius of curvature that was obtained using micromilling. Larger asperities can be obtained in the same way or, for example, through 3D-printing (*43*). For smaller scales, other well-established methods might be suitable, from laser ablation for asperities down to the micrometer scale (*44*) to micro- or nanolithography techniques at submicrometer scales (*45*).

In all cases, experimental challenges include the reproducibility of the shapes of asperities (for microcontact calibration to be relevant), the level of accuracy of the manufactured asperity heights with respect to their prescribed values and the potential misalignment between the two surfaces during the friction measurements [see (*26*) for a discussion of the latter two issues].

Many different types of frictional metainterfaces can be developed by using our design strategy, constituting a useful toolbox for designers of friction-based devices. By operating in optimized conditions, these devices should benefit from increased energy efficiency and lifetime. If further equipped with suitable sensors or actuators that bring relevant real-time changes to the topography, these metainterfaces could also become an asset in the development of smart systems (*46*) that incorporate functional solid contacts.

**Acknowledgments:** We thank A. Malthe-Sørenssen, K. Thøgersen, H.A Sveinsson, C. Oliver, M. Guibert, D. Bonamy and G. Pallares for discussions. We thank Zeyu Lin and Xin Qi for their help in developing the inversion scheme for friction laws defined by operating points. We thank D. Roux and N. Morgado for their help with the experimental device used in movie S1, and N. Morgado for the sketches in Fig. 1.

**Funding:** This work was funded by LABEX iMUST (grant ANR-10-LABX-0064) of Université de Lyon within the program "Investissements d'Avenir" (grant ANR-11-IDEX0007) operated by the French National Research Agency (ANR) (J.S.), the ANR through grant ANR-18-CE08-0011 (PROMETAF project) (J.S.), and the Institut Carnot Ingénierie@Lyon (PREGLISS project) (D.D.).

**Author contributions:** Conceptualization: J.S.; Methodology: A.A., J.S., D.D.; Investigation: A.A., J.S., E.D.; Visualization: A.A., J.S., D.D.; Writing – original draft: J.S.; Writing – review & editing: A.A., D.D., J.S.; Funding acquisition: J.S., D.D.

**Competing interests:** The authors declare that they have no competing interests.

**Data and materials availability:** All data are available in the main text or the supplementary materials.

**Supplementary Materials**

https://science.org/doi/10.1126/science.adk4234

Materials and Methods

Figs. S1 to S3

Tables S1 to S2

References (*47-51*)

Movie S1

# Supplementary Materials for

## Designing metainterfaces with specified friction laws

Antoine Aymard, Emilie Delplanque, Davy Dalmas, Julien Scheibert

Corresponding author: julien.scheibert@cnrs.fr

**The PDF file includes:**

    Materials and Methods
    Figures S1 to S3
    Tables S1 to S2
    References

**Other Supplementary Materials for this manuscript include the following:**

    Movie S1

**Materials and Methods**

Sample preparation

All PDMS samples are made of Sylgard 184, prepared according to the cross-linking protocol recommended in (*47*) (24 h at 25°C, demolding, and 24 h at 50°C). The samples are parallelepipedic blocks (thickness 7.2 mm, lateral sizes 20 and 20 mm), attached to a glass plate (see a typical picture in Fig. 1). The top surface is nominally flat, decorated by a square lattice of 8×8=64 spherical caps defining the surface asperities. The lattice period is 1.5 mm in both directions. The spherical caps are obtained by molding PDMS in an aluminum mold prepared with spherical holes, drilled with a sphere-ended tool of nominal radius of curvature 500 µm. Each hole has its individual depth, controlled with a precision of ±0.9 µm and setting the height $h_i$ of the summit of the corresponding spherical asperity. Four additional, deeper holes are drilled to generate four spherical asperities with large and identical altitudes, that will be used to align the contacting surfaces before mechanical testing (see Materials and Methods section "Mechanical testing"). The list of heights of the individual spheres differs for each surface, and is given below for each illustration example. The residual arithmetic mean areal roughness ($S_a$) of the surface of the spheres is 1.2 µm. The counter-surface is a smooth glass plate, cleaned first with acetone and then with distilled water.

Mechanical testing

The experiments are performed using the setup described in Fig. 13 of (*48*), placed in a clean room (temperature 20±1°C, humidity 50±10%). Forces are acquired at 10 Hz. The resolution of the normal force $P$ is ±5 mN, while it is ±(1 mN+0.01$F$) for the friction force $F$. Images of the real contact interface are taken as described in (*27*), using a high-resolution camera (Teledyne DALSA Genie Nano-GigE equipped with Qioptics optem fusion objective ; 4.6 to 5.0 µm/pixel, depending on the experiment). The contact area of each microcontact appears dark in the images (see Fig. 2A and insets of Figs. 3 to 5) and is measured by simple thresholding, the threshold value being determined individually using Otsu's method (*49*) applied to the image at maximum normal force.

The PDMS and glass nominally flat surfaces are first aligned to better than ±0.005°, thanks to four dedicated asperities with the very same large altitude, placed at the corners of the square lattice of asperities, and subsequently cut-off with a blade to enable subsequent testing of the metainterfaces.

Compression tests relating the normal force $P$ to the initial contact area $A_0$ are performed under controlled normal indentation, by steps of amplitude 1 µm and duration 40 s. One image is taken at the end of each step. Shearing tests enabling to measure the friction force $F$ and the associated (reduced) area $A_f$ are performed under constant normal force thanks to a feedback loop, and using a constant tangential velocity of 0.1 mm/s over a distance of 1 mm. During shear, images are taken at a rate of 8.3 frames per second. $F$ is measured as the peak of the curve of tangential force versus tangential displacement. $A_f$ is the contact area at the same instant.

Calibration of individual microcontacts

Single microcontact calibration is performed on samples featuring five identical asperities, distant by more than 7.1 mm to minimize elastic interactions between them (each contact's radius is always smaller than 400 µm). The individual contact areas and forces are taken as one fifth of the total contact areas and total forces. Those individual quantities are denoted $a$, $p$ and $f$ (lower case letters) to distinguish them from their macroscopic counterparts in metainterfaces, $A$, $P$ and $F$ (upper case letters). For each PDMS batch (see below), three normal-indentation–controlled compression tests have been performed (Fig. 2B). As in (50), we fit $\frac{a_0^{3/4}}{\sqrt{6}\pi^{5/4}R}$ as a function of $\frac{\pi^{1/4}p}{\sqrt{6}a_0^{3/4}}$, where $a_0$ (see Fig. 2A) and $p$ are, respectively, the contact area and normal force for a single microcontact under pure compression, and the curvature radius $R=526\pm5$ μm is measured from 48 different profilometry measurements on different individual spherical asperities. In this form, the data is found linear, indicating a good match with JKR's model of adhesive elastic spherical contacts (34). The fitted slope provides an estimate of the composite elastic modulus $E^*$. The intercept at origin provides an estimate of the effective adhesion energy. The latter is fitted to be about 9 mJ/m$^2$, a value so small that the data is very close to Hertz's model of adhesionless elastic spherical contact (34): for normal loads larger than 0.01 N, the relative difference in contact area between JKR's and Hertz's models is less than 2%. Thus, in our friction model, we used Hertz's model to relate $a_0$ to $p$, an assumption which proved to be sufficient for our purposes (see e.g. Fig. 2B). Two different batches of Sylgard 184 were used for this study, yielding either $E^*=1.36\pm0.04$ MPa (batch 1, curves in Figs. 3 and 4 and blue curve in Fig. 5) or $E^*=1.52\pm0.01$ MPa (batch 2, red curve in Fig. 5).

The shear-induced area-reduction ratio $B=0.85\pm0.01$ for batch 1 (resp. 0.92±0.01 for batch 2) is fitted on two (resp. one) curves $a_f$ versus $a_0$, using a linear function passing through the origin (Fig. 2C). $a_f$ is the microcontact area when the friction peak $f$ is reached (see Fig. 2A). Similarly, the friction strength $\sigma$ can be fitted on curves $f$ versus $a_f$, using a linear function passing through the origin (27) (see Fig. 2D for an example on batch 1).

Note on elastic interactions

Forces applied on a given surface asperity may affect the altitude of neighboring asperities due to long-range deformation of the underlying bulk solid. To assess the potential impact of such elastic interactions between microcontacts in our metainterfaces, we have compared our model of independent microcontacts to the model of (51), which explicitly incorporates elastic interactions. We used the geometrical parameters of our metainterfaces ($R$, in-plane positions of the spherical asperities, maximum individual contact radius), and looked at the relationships $A_0(P)$, $P(\delta)$ and $A_0(\delta)$, $\delta$ being the separation between the indenting rigid plane and the base plane that supports the asperities. We found that, while $P(\delta)$ and $A_0(\delta)$ are affected by interactions, the relationship of interest in our work, $A_0(P)$, is only negligibly affected (less than 2.3% rms difference between models with and without interactions, over all cases considered in the present work), in good agreement with the conclusions of (51). This is why, in our friction model, we could assume that microcontacts are independent.

Model parameter values

Model predictions use parameter values determined through dedicated experiments. For each batch, $R$, $E^*$ and $B$ are those extracted from the preliminary microcontact calibration (see Materials and Methods section "Calibration of individual microcontacts"). $\sigma$ is determined, for each metainterface, through friction experiments under normal forces so large that all asperities are involved in the contact. In this regime, not shown in Figs. 3 to 5, no new micro-contact is formed as the normal force is increased, and thus the friction behavior reduces to the non-linear sum over a constant number $N$ of Hertz-like micro-contacts. In practice, in the figures, to rescale the vertical axis, we used $\sigma=0.370\pm0.006$ MPa in Fig. 3 ; $\sigma=0.404\pm0.005$ MPa (resp. 0.358±0.006 MPa) for the blue data (resp. red data) in Fig. 4 ; $\sigma=0.450\pm0.004$ MPa (resp. 0.225±0.005 MPa) for the blue data (resp. red data) in Fig. 5.

Uncertainties

Each parameter has its own uncertainty, which contributes to the global uncertainty on the rescaled normal and friction forces. These uncertainties (see error bars in Figs. 3 to 5) are estimated, in a Monte-Carlo–like approach, as the standard deviation over 10000 evaluations of $\frac{F}{\sigma B}\left(\frac{P}{E^*}\right)$ for each $\frac{P}{E^*}$. In those calculations, the values of the forces and of the model parameters are drawn from gaussian distributions of mean (resp. standard deviation) the values (resp. error bars) provided in the present Materials and Methods.

Inversion scheme for friction laws defined by operating points

To design a metainterface with a friction law that passes through a series of operating points $\left(\frac{P_i}{E^*}, \frac{F_i}{\sigma B}\right)$ ($i \geq 1$), with $P_{i+1}>P_i$ and $F_{i+1}>F_i$, we propose to use the simplified type of topography sketched in fig. S1. For each operating point $i$, we seek the number $n_i$ of asperities needed to approach it from below, their common height $h_i$ (the $i$th height level), and the height $h_{i,a}$ of a single additional adjustment asperity to pass exactly through the operating point ($h_{i+1}<h_{i,a}<h_i$). In the context of the friction model described in the main text, the couple of equations describing the area and normal force as a function of the separation between the indenting rigid plane and the base plane that supports the asperities, $\delta$, is:

$$\begin{cases} \frac{F}{\sigma B} = A_0 = \pi R\left[\sum_{j=1}^{i}\left(n_j(h_j - \delta) + h_{j,a} - \delta\right) - l(h_{i,a} - \delta)\right] \\ \frac{P}{E^*} = \frac{4\sqrt{R}}{3}\left[\sum_{j=1}^{i}\left(n_j(h_j - \delta)^{3/2} + (h_{j,a} - \delta)^{3/2}\right) - l(h_{i,a} - \delta)^{3/2}\right] \end{cases} \quad (S1)$$

with $l=1$ when $h_{i,a}<\delta<h_i$ and $l=0$ when $h_{i+1}<\delta<h_{i,a}$.

$h_1$ represents the height of the highest asperities on the surface, and can be fixed at any arbitrary value. We then iteratively determine, for growing values of $i$, the values of $n_i$, $h_{i,a}$ and $h_{i+1}$. First, $n_i$ is the integer part of the solution of eqs. S1 while setting $l=1$, $F=F_i$ and $P=P_i$ (all other parameters are known). Note that if $n_i = 0$ already leads to a curve $\frac{F}{\sigma B}\left(\frac{P}{E^*}\right)$ passing above the $i$th operating point, the procedure is aborted because there is no solution to the considered list of operating points. Second, $h_{i,a}$ is the solution of eqs. S1 while setting $l=0$, $F=F_i$ and $P=P_i$. Third, $h_{i+1}$ is taken as the value of the solution $\delta$ of the same equations as for $h_{i,a}$.

In Fig. 3, the targeted operating points correspond to $F_1/\sigma B=0.84$, $F_2/\sigma B=1.68$ and $F_3/\sigma B=4.12$ mm², and $P_1=0.144$, $P_2=0.336$ and $P_3=0.649$ mm². Using the material parameters of batch 1, and setting the maximum asperity height to 270 μm, the design outputs are: $n_1=5$, $n_2=3$ and $n_3=47$ ; $h_1=270.0$, $h_2=180.7$ and $h_3=129.0$ μm ; $h_{1,a}=243.6$, $h_{2,a}=172.6$ and $h_{3,a}=115.2$ μm. A picture of this metainterface is shown in inset of Fig. 3, where the asperities at height level 1 (resp. 2 and 3) are circled in green (resp. blue and red).

Building a quasi-linear friction law

To design metainterfaces with linear-like friction laws, we use the following truncated exponential probability density function of asperity heights: $\Phi(h) = \frac{e^{-h/\lambda}}{\lambda(1-e^{-h_m/\lambda})}$ for $0 \leq h \leq h_m$, and $\Phi(h)=0$ elsewhere. $h_m$ is the maximum possible height, $\lambda$ is the scale parameter of the exponential, and there are $N$ asperities on the surface. If microcontacts follow Hertz's theory, in analogy with Greenwood & Williamson's model (*20*), the expected values of the friction and normal forces are:

$$\begin{cases} \frac{F(\delta)}{\sigma B} = A_0(\delta) = N\pi R \int_\delta^{h_m} \Phi(h)(h-\delta)dh \\ \frac{P(\delta)}{E^*} = N\frac{4}{3}\sqrt{R} \int_\delta^{h_m} \Phi(h)(h-\delta)^{3/2}dh \end{cases} \quad (S2)$$

with $\delta$ the separation between the indenting rigid plane and the base plane that supports the asperities. Both integrals have analytical solutions. When $h_m \to +\infty$, $\frac{F}{\sigma B}\left(\frac{P}{E^*}\right)$ is perfectly linear, in agreement with (20). When $h_m$ is finite, $\frac{F}{\sigma B}\left(\frac{P}{E^*}\right)$ remains essentially linear for large $P/E^*$, while larger non-linearities arise at small $P/E^*$ when $h_m/\lambda$ is smaller.

The asymptotic slope of $\frac{F}{\sigma B}\left(\frac{P}{E^*}\right)$ for large $P/E^*$ can be derived by linearizing both $F(\delta)/\sigma B$ and $P(\delta)/E^*$ around $\delta \to 0$ and combining them into a linear relationship $\frac{F}{\sigma B} \approx \alpha + m\frac{P}{E^*}$. The expression of the asymptotic slope is:

$$m = \sqrt{\frac{\pi R}{\lambda}} g\left(\frac{h_m}{\lambda}\right) \quad (S3)$$

where

$$g(x) = \frac{1-e^{-x}}{k(x)} \quad (S4)$$

with

$$k(x) = \text{erf}(\sqrt{x}) - \frac{2}{\sqrt{\pi}}\sqrt{x}e^{-x} \quad (S5)$$

and erf being the error function.

The intercept $\alpha$ is non vanishing when $h_m$ is finite, and reads $\alpha = \pi N R \lambda \Gamma\left(\frac{h_m}{\lambda}\right)$, where $\Gamma(x) = \frac{xe^{-x}}{1-e^{-x}}\left(\frac{4}{3\sqrt{\pi}}\sqrt{x}g(x) - 1\right)$.

For the illustration of Fig. 4, we targeted two different slopes $m$ ($m_t$ in Fig. 4). For each case, fixing $h_m$, the suitable value of $\lambda$ is then obtained by solving numerically the expression of $m$.

Once the parameters of the height distribution are chosen, we use the following method to create a discrete list of heights of $N$ asperities, whose collective contact behavior closely matches that of the continuous model of eqs. S2. The range of possible heights ($[0 ; h_m]$) is divided into elementary intervals of amplitude $\Delta h$, $[(i-1)\Delta h ; i\Delta h]$. The central value of each interval sets the corresponding height level, $h_i$. The number $n_i$ of asperities with this height is the closest integer to $N \int_{(i-1)\Delta h}^{i\Delta h} \Phi(h)dh$. In practice, we used $N=64$ and $\Delta h=6$ μm. The lists of $h_i$ and $n_i$ used for the two metainterfaces shown in Fig. 4 are given in table S1. In order to enable indentation beyond the point when all asperities are involved in a microcontact, even the smallest asperities must emerge from the base plane by a sufficient amount. Thus, an offset $h_0$ is added to the above-determined list of $h_i$ ($h_0$ is indicated in table S1). Note that such an offset has no expected impact on the resulting friction law. Also note that, as soon as all asperities are in contact, the friction behavior ceases to be linear-like, and becomes the non-linear sum over a constant number $N$ of Hertz-like micro-contacts, an example of which is the third branch in the friction law of Fig. 3.

Bounds on achievable friction coefficients
To evaluate the interval within which the friction coefficient can be targeted, we stick to the type of inversion used in Materials and Methods section "Specifying the friction coefficient at constant material pair": $R$ and $h_\mathrm{m}$ are fixed, and we look for the value of $\lambda$ that realizes the target value of the slope $m$ of the rescaled friction law $F/\sigma B$ vs $P/E^*$. $m$ can be rewritten for convenience as $m = \sqrt{\frac{\pi R}{h_\mathrm{m}}} \frac{\sqrt{x}(1-e^{-x})}{\mathrm{erf}(\sqrt{x}) + \left[\frac{1}{\sqrt{\pi x}} - 2\sqrt{\frac{x}{\pi}}\left(1 + \frac{1}{2x}\right)\right]e^{-x}}$, where the parameter $\lambda$ appears only through $x = h_\mathrm{m}/\lambda$. $m$ is a strictly increasing function of $x$, with the following two limits:

- when $x \to 0$, i.e. when $\lambda \to +\infty$ at fixed $h_\mathrm{m}$, then $m \to \frac{3\pi}{4}\sqrt{\frac{R}{h_\mathrm{m}}}$
- when $x \to +\infty$, i.e. when $\lambda \to 0$ at fixed $h_\mathrm{m}$, then $m \to +\infty$

Thus, in this framework, the theoretical interval in which the friction coefficient $\mu = \frac{\sigma B}{E^*} m$ can be targeted is $\left[\frac{3\pi \sigma B}{4 E^*}\sqrt{\frac{R}{h_\mathrm{m}}} \,;\, +\infty\right]$.

For instance, using $E^* = 1.36$ MPa, $\sigma = 0.358$ MPa, $B = 0.85$, $R = 526$ μm and $h_\mathrm{m} = 120$ μm (a realistic set of parameters in our experiments), the smallest targetable friction coefficient is 1.10.

Note that, in practice, arbitrarily large friction coefficients cannot be obtained, because those would require an arbitrarily small width of the distribution of asperity heights. In this case, the smallest imprecision in the manufacturing of the population of asperities will dominate the friction response. As a rule of thumb, we suggest that the minimum value of $\lambda$ for the distribution to remain reasonably exponential-like is ten times the typical manufacturing error on the asperities' heights. In our experiments, this error is about 0.9 μm, so that the minimum $\lambda$ would be 9 μm. With $h_\mathrm{m} = 120$ μm (a value used in our experiments, see table S1), $x = h_\mathrm{m}/\lambda \approx 13.33$. For such a large value of $x$, to a good approximation, the value of $m$ reduces to $\frac{\sigma B}{E^*}\sqrt{\frac{\pi R}{\lambda}}$. Using again a realistic set of parameters in our experiments ($E^* = 1.36$ MPa, $\sigma = 0.358$ MPa, $B = 0.85$, $R = 526$ μm), we find that the maximum targetable friction coefficient is 3.03.

We emphasize that the above-determined bounds on the achievable friction coefficient are specific to our experimental parameters when $\lambda$ is the only adjustable parameter of the design, and thus need to be evaluated again in any other context. For instance, letting $R$ be an adjustable design parameter could increase significantly the accessible range of $\mu$.

Building a bilinear friction law

The approach is similar to the one for quasi-linear friction laws (Materials and Methods section "Building a quasi-linear friction law"). We start with a continuous height distribution, $\Phi(h)$, defined as a modified exponential:

$$\Phi(h) = \begin{cases} \frac{1 - e^{-h_\mathrm{m}/\lambda} - u(e^{-h_\mathrm{c}/\lambda} - e^{-h_\mathrm{m}/\lambda})}{\lambda(1-e^{-h_\mathrm{c}/\lambda})(1-e^{-h_\mathrm{m}/\lambda})} e^{-\frac{h}{\lambda}} & \text{if } h \in [0\,;\, h_\mathrm{c}] \\ \frac{u}{\lambda(1-e^{-h_\mathrm{m}/\lambda})} e^{-\frac{h}{\lambda}} & \text{if } h \in [h_\mathrm{c}\,;\, h_\mathrm{m}] \\ 0 & \text{if } h_\mathrm{m} < h \end{cases} \quad (S6)$$

with $\lambda$ the scale parameter of the exponential, $h_\mathrm{m}$ the maximum height, $h_\mathrm{c}$ the cross-over height separating the two branches in the distribution and $u$ a positive scalar parameter useful to tune the respective weights of the two branches. $F(\delta)/\sigma B$ and $P(\delta)/E^*$, with $\delta$ the separation between the indenting rigid plane and the base plane that supports the asperities, are obtained by inserting $\Phi(h)$ (eq. S6) into eqs. S2. For $u=1$, both weights are equal, $\Phi(h)$ corresponds to the unweighted truncated exponential treated in Materials and Methods section "Building a quasi-linear friction law", and the friction law is quasi-linear. For smaller $u$, the two weights are different, yielding a second quasi-linear branch in the friction law (see Fig. 5). The cross-over normal force, $P_\mathrm{c}$, is the value of $P$ when $\delta = h_\mathrm{c}$.

For each quasi-linear branch, we calculate the asymptotic slope of $\frac{F}{\sigma B}\left(\frac{P}{E^*}\right)$ for large $P/E^*$, $m_1$ for the first branch $P \leq P_\mathrm{c}$ ($\delta \geq h_\mathrm{c}$) and $m_2$ for the second branch $P > P_\mathrm{c}$ ($\delta < h_\mathrm{c}$). To design metainterfaces with prescribed $m_1$, $m_2$ and $P_c/E^*$, we first fix $h_m=120$ µm and solve numerically for $\lambda$ using the equation of $m_1$, which does not involve $u$ nor $h_c$:

$$m_1 = \sqrt{\frac{\pi R}{\lambda}} g\left(\frac{h_m}{\lambda}\right) \tag{S7}$$

with $g$ defined in eq. S4.

$u$ and $h_c$ are then obtained simultaneously through numerical solution of the coupled equations for $m_2$ and $P_c/E^*$:

$$\begin{cases} m_2 = \sqrt{\frac{\pi R}{\lambda}} \frac{\alpha_1 - \alpha_2 e^{-\frac{h_m}{\lambda}} - (\alpha_1 - \alpha_2) e^{-\frac{h_c}{\lambda}}}{(\alpha_1 - \alpha_2) k\left(\frac{h_c}{\lambda}\right) + \alpha_2 k\left(\frac{h_m}{\lambda}\right)} \\ \frac{P_c}{E^*} = \frac{4}{3} N \sqrt{R} \alpha_2 \left[\frac{3}{4} \sqrt{\pi} \lambda^{5/2} e^{-\frac{h_c}{\lambda}} \mathrm{erf}\left(\sqrt{\frac{h_m - h_c}{\lambda}}\right) + \lambda e^{-\frac{h_m}{\lambda}} \frac{\sqrt{h_m - h_c}}{2} (2h_c - 3\lambda - 2h_m)\right] \end{cases} \tag{S8}$$

with $k$ the function defined in eq. S5, $\alpha_2 = \frac{u}{\lambda(1 - e^{-h_m/\lambda})}$ and $\alpha_1 = \frac{1 - \lambda \alpha_2 (e^{-h_c/\lambda} - e^{-h_m/\lambda})}{\lambda(1 - e^{-h_c/\lambda})}$.

For the two examples shown in Fig. 5, we targeted a common couple of slopes, $m_1=7.15$ and $m_2=12.52$, but two different cross-over forces $P_c/E^*$ (denoted $P_{c,t}/E^*$ in Fig. 5), either 0.0380 or 0.0737 mm$^2$. From the data in Fig. 5, the actual $P_c$ (denoted $P_{c,f}/E^*$ in Fig. 5) is estimated as the average normal load between the last point that is well-captured by the linear fit of the first branch and the first point deviating from it: $P_{c,f}/E^*=0.046\pm0.005$ and $0.098\pm0.005$ mm$^2$ for the red and blue curves, respectively.

For each metainterface, the discrete list of asperity heights, $h_i$, that offers a good quantitative agreement with the prediction from the continuous model of eqs. S2, is obtained as follows. One first calculates, for $N=64$, the numbers of asperities in regime 1 ($h>h_c$) and regime 2 ($h<h_c$), $N_1$ and $N_2$ respectively. For each regime, the height range ($[h_c\,;\,h_m]$ and $[0\,;\,h_c]$ for regimes 1 and 2, respectively) is divided into a number of intervals equal to $N_1$ or $N_2$, respectively. The width of each interval is calculated such that the probability that $h$ belongs to the interval $[b_i\,;\,b_{i+1}]$, $\int_{b_i}^{b_{i+1}} \Phi(h) dh$, is equal to $1/(N_1+N_2)$. Then, a single asperity is placed at the center of each so-determined interval.

The lists of $h_i$ underlying the curves in Fig. 5 are provided in table S2 (the values include an offset $h_0=50$ µm, for the same reasons as those mentioned at the end of Materials and Methods section "Building a quasi-linear friction law").

Applying the design strategy to the sliding friction force

In our shearing calibration experiments, one can define two different friction forces, either the peak (static) friction force at the onset of sliding (as done in Fig. 2) or the plateau (sliding) friction force when macroscopic motion has settled (in practice, $f$ is then defined as the average tangential force for tangential displacements in the range $[100 - 200]$ µm after the static friction peak ; $a_f$ is defined analogously). We find that both friction forces obey the same behavior laws ($f=\sigma a_f$ and $a_f=Ba_0$) but that the calibrated values of $\sigma$ and $B$ are slightly modified. Such modifications only affect the rescaling factor of the macroscale friction force, but will not affect the predicted curves $\frac{F}{\sigma B}\left(\frac{P}{E^*}\right)$ shown in Figs. 3 to 5. And indeed, for all three examples, we have found that the agreement between the various targeted frictional features and the corresponding measurements is, quantitatively, similarly good for the static friction force (shown in Figs. 3 to 5) and sliding friction forces.

Suitability of an elastic asperity model

While elastomers like PDMS remain elastic up to very large strains, other materials are likely to experience plastic deformation when involved in a rough contact. The question thus arises of the limits of validity of elastic asperity models to describe such elasto-plastic contacts. It is addressed in (20), where a plasticity index is defined as $\psi=(E^*/H).(\lambda/R)^{1/2}$, with $E^*$ and $H$ the composite elastic modulus and hardness of the material, respectively, $R$ the curvature radius of the asperities and $\lambda$ the standard deviation of their distribution of heights. The authors of (20) argue that an elastic asperity model is suitable if $\psi<0.6$, which leads us to propose the following interface design steps:

1. choice of the material, which imposes $E^*$ and $H$.
2. inversion of an elastic asperity model to identify suitable couples of geometrical parameters, $R$ and $\lambda$, such that the specifications on the friction law are met.
3. check whether one of those couples satisfies the criterion $\psi<0.6$.
   - if yes, then a design solution is found
   - if not, the elastic asperity model cannot be used for this couple of $E^*$ and $H$. Then, go back to step 1 and consider another material with a smaller ratio $E^*/H$. Or, go back to step 2 and replace the elastic model with another asperity model explicitly accounting for plasticity (40, 41).

Effect of a misalignment on the friction law

The design strategy provides, from a given set of specifications on the target friction law, a suitable list of asperities' heights, $h_i$. This list of heights is the relevant one as long as the mean planes of the two contacting solids are perfectly parallel. In experiments however, a non-vanishing residual angle between both planes exists. In our experiments, this angle remains in the interval [-0.005° ; +0.005°], thanks to a specific alignment procedure (see Materials and Methods section "Mechanical testing"). To assess the potential effect on the experimental friction law of such imperfect conditions, we perform numerical simulations using the very same asperity-based friction model as that used in the rest of this work (see description in the third paragraph of section "Applying the strategy to centimetric elastomer-on-glass interfaces").

To account for an angle $\phi$ around both in-plane axes $x$ and $y$ simultaneously, we modify the prescribed list of heights by adding a contribution $\Delta h_i = h_i + x_i.\tan\phi + y_i.\tan\phi$, with $x_i$ and $y_i$ the coordinates of the $i$th asperity. The origin $x=y=0$ is taken at one corner of the square lattice of asperities. We then apply the friction model to the updated list of heights, $h_i + \Delta h_i$.

As examples, we consider the substrates associated to the red and blue friction laws in Fig. 4, and we vary $\phi$ from -0.1° to +0.1° (yielding a maximum height modification of about 37 μm in our experiments), by steps of 0.0001°. We observe that, for each value of $\phi$, the simulated rescaled friction law, $F/\sigma B$ versus $P/E^*$, keeps a quasi-linear shape, so that the effect of the angle can be assessed using the slope of the law, as fitted in the same way as in Fig. 4. The fitted slope is denoted $m_\phi$, while the target slope for $\phi=0$ is denoted $m_0$. We quantify the discrepancy between the target slope and the one for misaligned surfaces as the standard deviation (SD) of $m_\phi - m_0$ over all simulations with an angle in the interval [-$\phi$ ; +$\phi$]. Note that this quantity is a good estimate of the uncertainty on the experimental slope when the alignment precision is $\pm\phi$.

Figure S2 shows the simulated evolution of the relative discrepancy between the target slope and that for a misaligned interface, SD($m_\phi-m_0$)/$m_0$, as a function of |$\phi$|, for both samples in Fig. 4 (solid lines). Note that, for each sample, the shape of the curve is strongly related to the precise in-plane locations of the various asperities: simulations using the very same list of heights but shuffled in-plane locations yield very different curves (dashed lines). Those large differences are presumably due to the small number of asperities along the interface (64 asperities). The curves obtained for both samples are in separate parts of the plot, presumably due to the two different values of λ, the scale parameter of the exponential in the distribution of heights for both samples.

Figure S2 shows that, for the samples of Fig. 4, the typical discrepancy on the slope of the rescaled friction law due to misalignment is less than 2.1% when $\phi$ is smaller than 0.1°. Note that, for our uncertainty on the experimental alignment ($\pm 0.005°$), this discrepancy remains always smaller than about 0.2% (see dots in fig. S2). Such a small value has been an asset contributing to our successful quantitative comparisons between target and experimental friction laws.

Effect on the friction law of manufacturing errors on the asperities' heights
Once a list of heights, $h_i$, has been identified as a solution to a set of specifications on the friction law, a corresponding sample is manufactured. The manufacturing process generally results in an imperfect list of heights, which can be described as $h_i + dh_i$, where $dh_i$ is the error on the height of the $i$th asperity. To investigate the effect of those errors on the friction law, we perform numerical simulations in which we use our asperity-based friction model to calculate the friction law resulting from the perturbed list of height, $h_i + dh_i$. All other model parameters are kept unchanged. Two reference lists $h_i$ are considered: those underlying the two quasi-linear friction laws in Fig. 4.
Each list $dh_i$ is randomly drawn from a centered Gaussian distribution of standard deviation SD($dh$). SD($dh$) is varied from 0 to 3 µm, by steps of 0.1 µm. For each SD($dh$), 10000 simulations are performed, each with a different draw, and the slope $m$ of the rescaled friction law is fitted in the same way as in Fig. 4.
Figure S3 shows the evolution, as a function of SD($dh$), of the average slope <$m$> (over the 10000 simulations) divided by $m_0$, the target slope in the absence of manufacturing errors. We find that <$m$>/$m_0$ remains very close to 1, indicating that the perturbation of the topography has almost no effect on the average friction law. However, the standard deviation on $m$, SD($m$), has a clear, linear-like dependence on SD($dh$), as is visible through the amplitude of the error bars ($\pm$SD($m$)/$m_0$). Thus, for a particular realization of a perturbed topography, the experimental slope of the friction law has a larger probability to deviate from the target slope when SD($dh$) in larger. Figure S3 indicates that the typical deviation is smaller than 1.5% in all our simulations. In our experiments, in which SD($dh$)=0.9 µm (Materials and Methods section "Samples preparation"), the typical deviation of the slope is smaller than 0.5%. Such a small value has contributed to the success of our quantitative comparison between target and experimental friction laws.

**Fig. S1.**

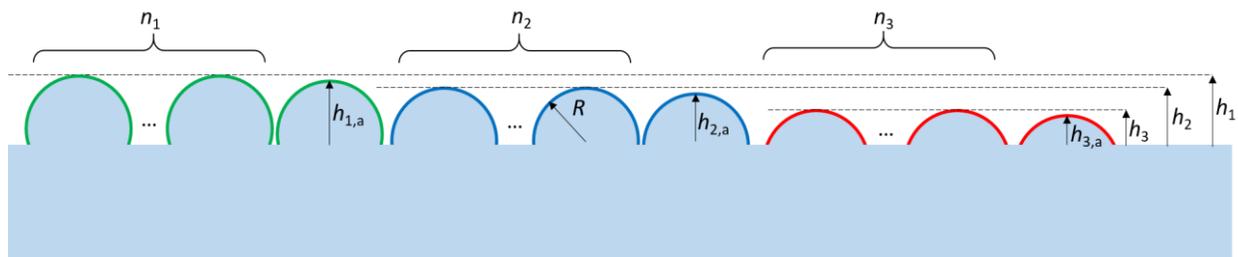

**Fig. S1. Sketch of the topography used to target three operating points.** $N = n_1 + n_2 + n_3 + 3$ spherical asperities of common radius $R$ have different heights. $n_i$ asperities of height $h_i$ complemented by an additional asperity of height $h_{i,a}$ enable reaching operating point $i$. The three colors of the asperities are the same as that of the corresponding operating points in Fig. 3.

**Fig. S2.**

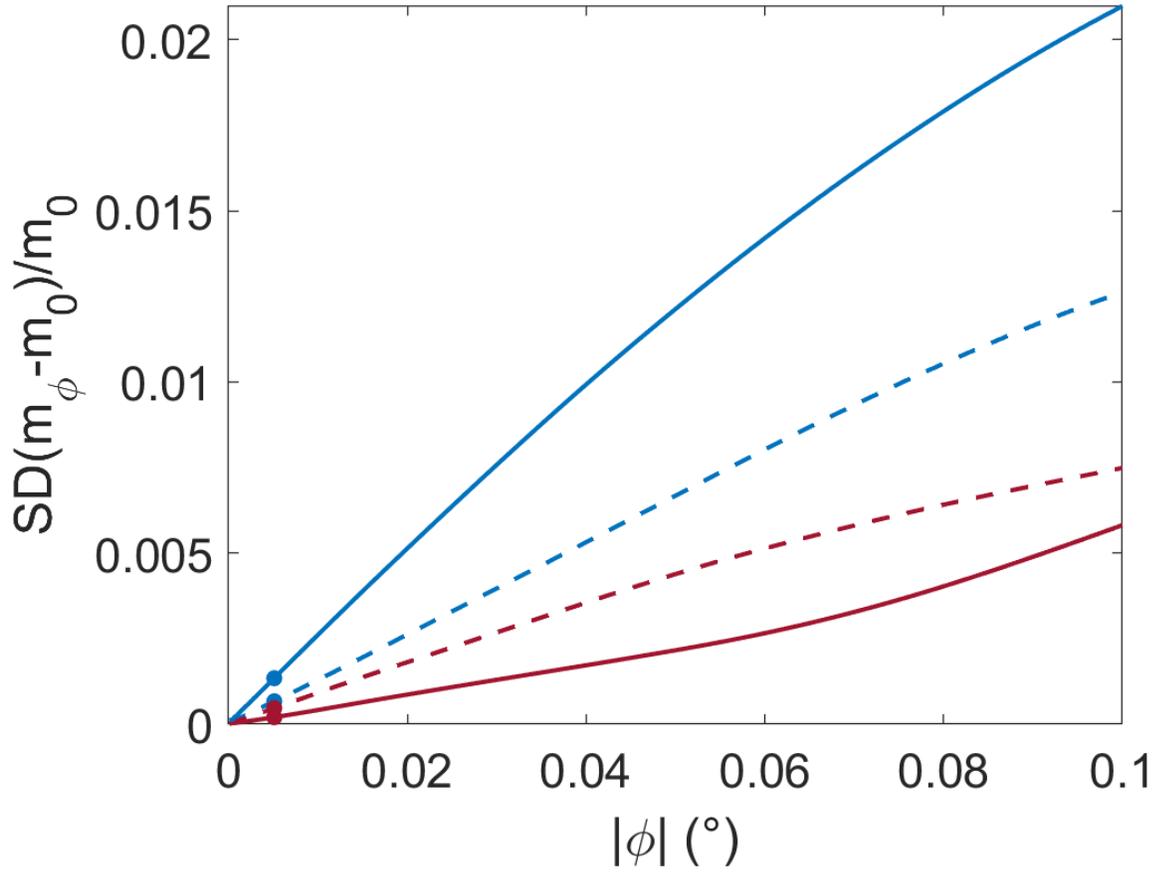

**Fig. S2. Effect of a misalignment of the contacting surfaces on the friction law.** Results of simulations based on the asperity-based friction model, applied to the lists of heights of both samples shown in Fig. 4 (same color code), modified by the introduction of an angle $\phi$ around both axes $x$ and $y$ simultaneously (for details, see Materials and Methods section "Effect of a misalignment on the friction law"). Solid lines: evolutions of the standard deviation (SD, over all simulations with an angle in [-$\phi$ ; +$\phi$]) of the difference between the fitted slopes of the rescaled friction law, with and without misalignment ($m_\phi$ and $m_0$ respectively), rescaled by $m_0$, as a function of the absolute value of $\phi$. Dashed lines: similar curves for simulations using the same list of heights but shuffled lateral positions of the asperities. Dots correspond to the experimental precision on the alignment ($\pm$ 0.005°).

**Fig. S3.**

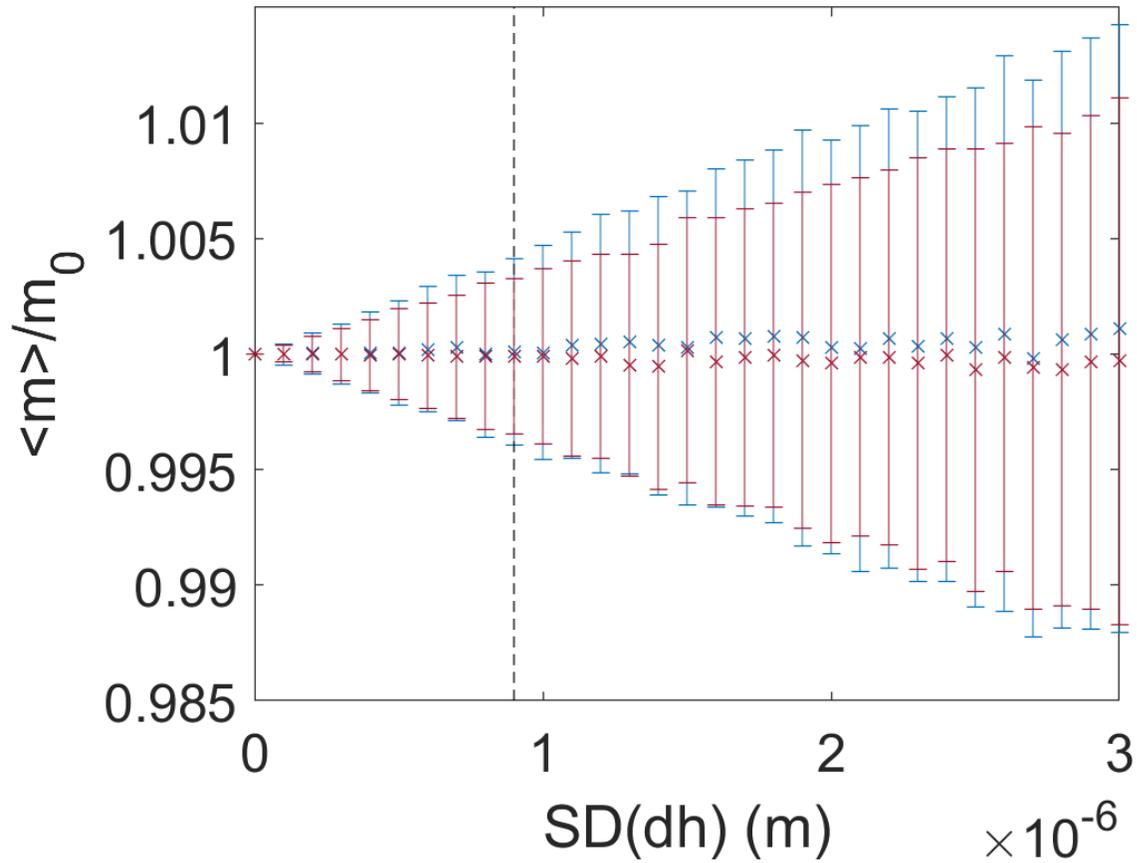

**Fig. S3. Effect on the friction law of manufacturing errors on the asperities' heights.** Results of simulations based on the asperity-based friction model, applied to the lists of heights of both samples shown in Fig. 4 (same color code), after adding to each asperity $i$ a height perturbation $dh_i$ (for details, see Materials and Methods section "Effect on the friction law of manufacturing errors on the asperities' heights"). Mean slope of the rescaled friction law over 10000 realizations of the list of $dh_i$, divided by the target slope $m_0$ (for a vanishing perturbation), as a function of the standard deviation of the perturbation, SD($dh$). Error bars correspond to $\pm$SD($m$)/$m_0$. The vertical dashed line indicates SD($dh$)=0.9 µm, which is the precision on the asperities' height in our experiments.

**Table S1.**

| $h_i$ (μm) | 3 | 9 | 15 | 21 | 27 | 33 | 39 | 45 | 51 | 57 | 63 | 69 | 75 | 81 | 87 | 93 | 99 | 105 | 111 | 117 |
|---|---|---|---|---|---|---|---|---|---|---|---|---|---|---|---|---|---|---|---|---|
| $n_i$ (blue curve, $\lambda$=20 μm, $h_\mathrm{m}$=74 μm, $h_0$=80 μm) | 17 | 12 | 9 | 7 | 5 | 4 | 3 | 2 | 2 | 1 | 1 | 1 | 0 | 0 | 0 | 0 | 0 | 0 | 0 | 0 |
| $n_i$ (red curve, $\lambda$=40 μm, $h_\mathrm{m}$=120 μm, $h_0$=84 μm) | 9 | 8 | 7 | 6 | 5 | 4 | 4 | 3 | 3 | 2 | 2 | 2 | 2 | 1 | 1 | 1 | 1 | 1 | 1 | 1 |

**Table S1. List of prescribed heights for metainterfaces with a quasi-linear friction law**. The colors refer to the curves in Fig. 4. Notations refer to Materials and Methods section "Building a quasi-linear friction law".

**Table S2.**

| Asperity | 1 | 2 | 3 | 4 | 5 | 6 | 7 | 8 | 9 | 10 | 11 | 12 | 13 | 14 | 15 | 16 |
|---|---|---|---|---|---|---|---|---|---|---|---|---|---|---|---|---|
| $h_i+h_0$ (μm) (blue curve) | 53.8 | 60.6 | 60.3 | 55.3 | 50.1 | 61.0 | 52.0 | 101.5 | 59.8 | 57.3 | 51.8 | 60.8 | 54.0 | 57.7 | 58.2 | 50.5 |
| $h_i+h_0$ (μm) (red curve) | 63.3 | 62.4 | 58.7 | 60.9 | 56.0 | 57.6 | 55.5 | 54.3 | 60.6 | 55.8 | 56.3 | 55.0 | 94.5 | 64.9 | 58.9 | 59.8 |

| Asperity | 17 | 18 | 19 | 20 | 21 | 22 | 23 | 24 | 25 | 26 | 27 | 28 | 29 | 30 | 31 | 32 |
|---|---|---|---|---|---|---|---|---|---|---|---|---|---|---|---|---|
| $h_i+h_0$ (μm) (blue curve) | 59.1 | 53.2 | 53.4 | 50.7 | 57.0 | 51.3 | 54.6 | 58.8 | 54.4 | 69.1 | 87.2 | 52.2 | 53.6 | 62.6 | 56.8 | 56.1 |
| $h_i+h_0$ (μm) (red curve) | 62.6 | 61.2 | 54.5 | 50.8 | 56.8 | 61.5 | 52.9 | 52.2 | 59.2 | 61.8 | 57.3 | 56.5 | 50.4 | 58.3 | 50.6 | 65.2 |

| Asperity | 33 | 34 | 35 | 36 | 37 | 38 | 39 | 40 | 41 | 42 | 43 | 44 | 45 | 46 | 47 | 48 |
|---|---|---|---|---|---|---|---|---|---|---|---|---|---|---|---|---|
| $h_i+h_0$ (μm) (blue curve) | 51.6 | 81.7 | 152.4 | 51.5 | 52.8 | 50.9 | 55.5 | 126.2 | 54.9 | 55.9 | 56.4 | 52.6 | 111.7 | 60.1 | 59.3 | 57.5 |
| $h_i+h_0$ (μm) (red curve) | 54.8 | 50.2 | 64.3 | 110.9 | 62.1 | 51.0 | 51.3 | 83.6 | 52.6 | 75.2 | 53.6 | 64.5 | 54.0 | 59.5 | 60.0 | 53.8 |

| Asperity | 49 | 50 | 51 | 52 | 53 | 54 | 55 | 56 | 57 | 58 | 59 | 60 | 61 | 62 | 63 | 64 |
|---|---|---|---|---|---|---|---|---|---|---|---|---|---|---|---|---|
| $h_i+h_0$ (μm) (blue curve) | 93.6 | 50.3 | 59.6 | 57.9 | 55.7 | 56.6 | 52.4 | 51.1 | 77.0 | 55.1 | 54.2 | 53.0 | 58.4 | 58.6 | 65.7 | 72.8 |
| $h_i+h_0$ (μm) (red curve) | 145.4 | 51.9 | 60.3 | 63.9 | 51.7 | 63.0 | 52.4 | 51.5 | 55.3 | 53.3 | 57.8 | 57.0 | 63.6 | 53.1 | 68.4 | 58.1 |

**Table S2. List of prescribed heights for metainterfaces with a bilinear friction law.** The colors refer to the curves in Fig. 5. Notations refer to Materials and Methods section "Building a bilinear friction law".

**Movie S1.**

**Movie S1.** Qualitative illustrative experiment showing that two metainterfaces may have either similar or different frictional behaviour, depending on the range of normal force applied. The metainterfaces were fabricated from the same two molds used for Fig. 5. The low normal force corresponds to the weight of the bare samples, about 5 g. The large normal force corresponds to an additional weight of about 10 g (rectangular polymer plate placed on top of each sample). For each normal force, both samples are driven by the same tangential displacement $\Delta$, through loading springs. When the samples are sliding (non-vanishing sliding distances $S_R$ and $S_B$), the spring lengths ($L_R$ and $L_B$) enable visualization of whether the friction forces are similar or different.